\documentclass[10pt]{iopart}
\usepackage{graphicx,hyperref,amsfonts,amssymb,mathrsfs}
\usepackage{epstopdf,color,bm,multirow,rotating,soul,comment}
\usepackage[caption=false]{subfig}
\usepackage[absolute,overlay]{textpos}

\begin{document}

\title[Superefficient cascade multiresonator quantum memory]{Superefficient cascade multiresonator quantum memory}
\author{N S  Perminov$^{1,2}$, D Yu  Tarankova$^{3}$, and S A Moiseev$^{1,2,*}$}

\address{$^{1}$ Kazan Quantum Center, Kazan National Research Technical University n.a. A.N.Tupolev-KAI, 10 K. Marx, Kazan 420111, Russia}
\address{$^{2}$ Zavoisky Physical-Technical Institute, Kazan Scientific Center of the Russian Academy of Sciences, 10/7 Sibirsky Tract, Kazan 420029, Russia}
\address{$^{3}$ Institute of Radio-Electronics and Telecommunications, Kazan National Research Technical University n.a. A.N.Tupolev-KAI, 10 K. Marx, Kazan 420111, Russia}
\ead{s.a.moiseev@kazanqc.org}

\vspace{10pt}
\begin{indented}
\item[] \today
\end{indented}

\begin{abstract}
We propose a cascade scheme of a superefficient broadband quantum memory consisting of four high-Q ring resonators forming a controllable frequency comb and interacting with long-lived spin systems and with a common waveguide.
Using the transfer function giving extended matching conditions, the optimization of all spectroscopic parameters of the system for quantum memory in the resonator is carried out.
It was shown that our quantum memory scheme does not impose large restrictions on the parameters of losses in resonators and allows to achieve super high efficiency $\sim99.99\%$ in a wide frequency range.
\end{abstract}

\vspace{2pc}
\maketitle
\ioptwocol

\section{Introduction}
The creation of an effective quantum interface (QI) and long-lived quantum memory (QM) is of critical importance for quantum information technologies \cite{Hammerer2010,Kurizki2015}.
In practical implementation of long-lived multiqubit QM, it is required an implementation of sufficiently strong and reversible interaction of light/microwave qubits with many long-lived information carriers \cite{Roy2017}, in particular with NV-centers in diamond \cite{Jiang2009} and rare-earth ions in inorganic crystals \cite{Zhong2015}.
The best realization of this approach provided quantum storage with efficiency up to 92 \% \cite{Cho2016,Hsiao2018}, while full-fledged multifunctional quantum computer requires at least 99.9 \%.
The solution of the problem of high efficiency in operation of QI is possible due to the using the opportunities of rich dynamics in many-particle systems \cite{Hartmann2008,Hur2016,Noh2017}.
Herein, there is a basic problem in a construction of the multi-particle systems demonstrating controlled perfect time-reversible dynamics. 

The considerable improvement of high-Q resonators \cite{Gorodetsky1999,Vahala2003,Kobe2017,Toth2017,Megrant2012} and their integration into multiresonator structures \cite{Armani2003,Liu2018,Xie2018,Flurin2015,Pfaff2017,Sirois2017} makes them interesting for use in broadband efficient optical and microwave QIs  \cite{EMoiseev2017,Moiseev_2017_PRA,Moiseev2018} in which the resonators can have a specified periodic frequencies. 
Such multi-resonator schemes demonstrate the possibility of a significant increase in the operating spectral range of the QI.
Moreover high quality of the resonators makes it possible to considerable enhance the constant coupling with light signals and resonant atomic ensembles herewith the broadband system of resonators allows reducing the effects of relaxation and decoherence due to the transition to faster storage processes.
These properties promise getting higher QM efficiency and using these systems in circuits of the universal quantum computers \cite{Kurizki2015,Perminov2017superefficient,Perminov2017spectral,Kockum2018}.

In this paper we show that combining the system of high-Q ring resonators with long-lived spin ensembles could be a realible cascade approach to obtain superefficient long-lived QM in practically realistic conditions. 
We introduce the term "superefficient" by indicating the possibility to increase the quantum efficiency closely to 100 \% due to using the system of high-Q resonators providing an efficient broadband QI between free propagating signal photons and long-lived spin ensemble situated in these resonators.

\begin{figure}[t]
	\includegraphics[width = 0.45\textwidth]{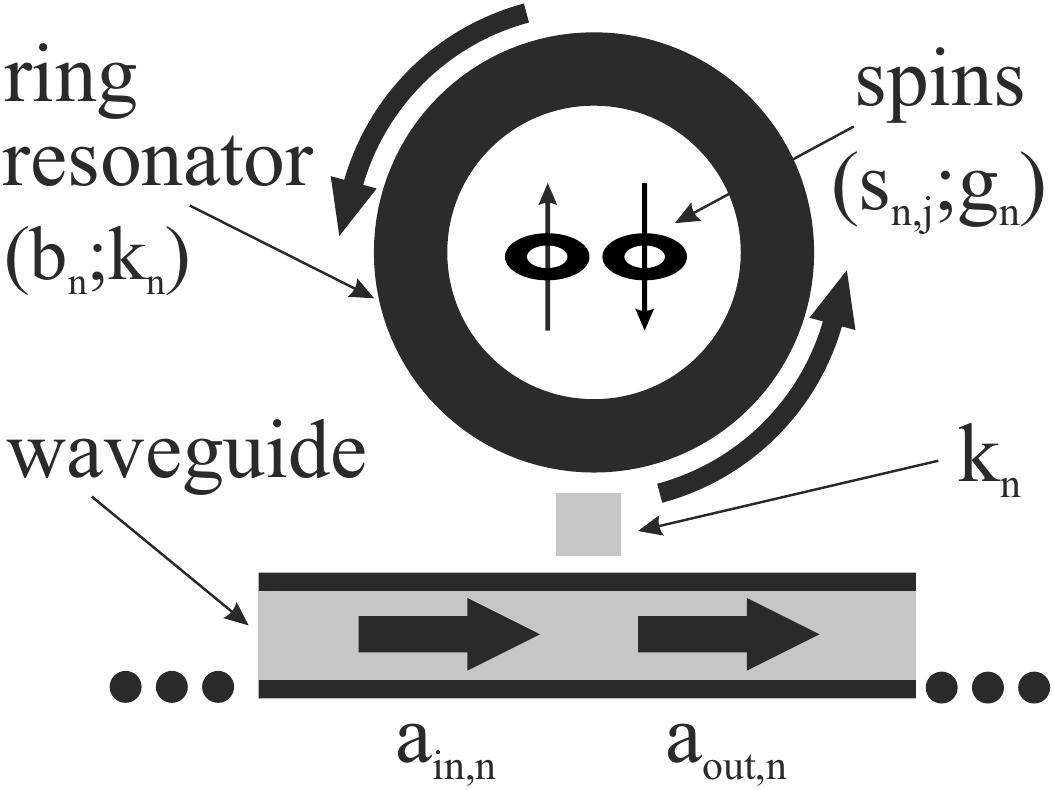}
	\caption{Elementary building block of the scheme of controllable frequency comb quantum interface based on ring resonators.}
	\label{Scheme}
\end{figure}

The proposed cascade scheme consists of four high-Q ring resonators forming a controllable frequency comb (CFC) and interacting with long-lived resonant spin systems and with a common broadband waveguide (Fig.~\ref{Scheme}).
The cascade regime of the quantum scheme is realized in two stages.
Firstly the input signal is transferred to the ring resonators from the external waveguide and then the signal field is stored in the system of long-lived spin ensembles.
A small number of resonators greatly facilitates the spectral-topological optimization of parameters of the CFC QI scheme \cite{Perminov2017spectral}.

The simple model of spectral optimization proposed below was based for the fitting the transfer function (TF) of the studied system describing to the spectral characteristics of the quantum storage process.
It was shown that the long-lived CFC QM scheme does not impose large restrictions on the parameters of losses in resonators.
By using this optimization method we have found the optimal values of all free parameters of the CFC QI scheme and show the possibility of achieving the super high efficiency $\sim99.99\%$ in a wide frequency range.
Finally we discuss possible experimental implementations of the proposed scheme and it's potential application.

\section{Physical background}
The initial idea of the scheme under consideration is based on the photon echo QM approach \cite{Moiseev2001,Moiseev2004,Tittel2009} in a variant using resonant atoms with a periodic spectral structure of the inhomogeneous broadening of the absorption line, which is known as the AFC protocol \cite{Riedmatten2008,Akhmedzhanov2016}, and also on the realization of this approach in the optimal resonator \cite{Moiseev2010,Afzelius2010}.
This approach has been recently extended to systems of several resonators \cite{EMoiseev2017,Moiseev_2017_PRA,Perminov2017superefficient}.

For CFC QM, we analyze the dynamics of $N$ resonators and resonant two-level spin ensembles by using the quantum optics approach to the description of light in an open cavity \cite{Walls}.
Spin ensembles exist in each ring resonator and resonant frequencies of the spins close to the frequency of its ring cavity mode.  
Herein resonant transition of each spin ensemble is characterized by inhomogeneous broadening and spectrum of the input signal field covers the spectrum of ring frequency combs. 
We also assume that all two-level spin ensembles are prepared in the ground states $|G_s\rangle=\prod_{j=1}^N|g_j\rangle$ before signal pulses described by the initial state $|\psi_{in,1}\rangle$ is launched into the studied system. 
In the framework of this approach, we get the system of Langevin-Heisenberg equations for the spin modes $s_{n,j}(t)$, the resonator field modes $b_n(t)$ and the input-output mode of the common waveguide $a_{in,out,n}(t)$:
\begin{eqnarray}\label{eq}
\nonumber \left[\partial_{t}+i(\tilde{\Delta}_n+\delta_{n,j})\right]s_{n,j}(t)+ig_nb_n(t)=0, &\\
\nonumber \left[\partial_t +i\Delta_n'+\frac{\kappa_n}{2}\right]b_n(t)+
i \sum_jg_ns_{n,j}(t)=\\ \nonumber k_n^{\frac{1}{2}}a_{in,n}(t)+{\gamma}_n^{\frac{1}{2}}F_n(t),& \\
\nonumber a_{in,n}(t)-a_{out,n}(t) =k_n^{\frac{1}{2}}b_n(t), &\\
a_{in,n+1}(t)=e^{i\varphi_n}a_{out,n}(t+(z_{n+1}-z_n)/c),
\end{eqnarray}
where $\varphi_n=\omega_0 (z_{n+1}-z_n)/c$, $z_n$ is the spatial position of the $n$-th resonator along the waveguide, $\omega_0$ is a central frequency, 
$\delta_{n,j} $ is the frequency detuning of the $j$-th spin in the $n$-th spin system (reckoned from its line center) with detuning $\tilde{\Delta}_n$ ($j\in\{1,...,N_n\}$), $\{g_n,\delta_n=1/T_{2,n}^*\}$ is the field interaction constant in the $n$-th resonator with a single spin and the inhomogeneous linewidth of the $n$-th spin system (further we shall assume that the spin systems in the resonators are identical and $\delta_n=\tilde{\delta}$),
$\{\tilde{\Delta}_n=\Delta(n-\textrm{sgn}(n)/2),\Delta_n\}$ is the frequency detuning of centers of lines of spin systems and resonators, $n\in\{-N/2,...,N/2\}\backslash\{0\}$ (in this paper we consider the case of an even $N$); $\Delta_n'=\Delta_n-i\gamma_n$, where $\Delta_n$ and $\gamma_n$ are spectral detuning and decay constant of $n$-th cavity mode with corresponding Langevin forces $F_n(t) $\cite{Scully}.

Assuming the number of spins in each resonator is a quite large $N_n\gg1$, the discrete value $\delta_{n,j}$ is replaced by the frequency detuning $\delta$ with continuous distribution of the Lorentzian line shape $G_n(\delta)=\pi^{-1}{\delta}_n[\delta^2+{\delta}_n^2]^{-1}$.
Next we will use dimensionless units for all the parameters in (\ref{sol_eq}), which is equivalent to relating them to a certain unit of frequencies, assuming without loss of generality $\Delta=1$.
We also ignored the Langevin forces $F_n(t)$ \cite{Walls} in the equation (\ref{eq}), focusing only on the searching for the quantum efficiency in the CFC scheme studied and by the case of ultimately weak decay constants  $\gamma_{1,...,N} \ll\kappa_1$  and $\gamma_{1,...,N} \tau\ll 1$ (where $\tau$ is a typical time of the studied processes).

\section{Recording/reading stage}	
By taking the Fourier transform of (\ref{eq}), we obtain the system of algebraic equations for recording stage
\begin{eqnarray}\label{eq_four}
\nonumber i(\tilde{\Delta}_n+\delta_{n,j}-\omega)s_{n,j}+ig_nb_n=0, &\\
\nonumber \left[i(\Delta_n'-\omega)+\frac{\kappa_n}{2}\right]b_n+i \sum_jg_ns_{n,j}=k_n^{\frac{1}{2}}a_{in,n}, &\\
\nonumber a_{in,n}-a_{out,n}(\omega)=k_n^{\frac{1}{2}}b_n,&\\
a_{in,n+1}=e^{i(\varphi_n+\omega (z_{n+1}-z_n)/c)}a_{out,n}(\omega), 
\end{eqnarray}
where for all the field and spin operators the Fourier transform is defined as $u(t)=[2\pi]^{-\frac{1}{2}}\int d\omega u(\omega) e^{-i\omega t}$,
where $\omega$ is the frequency counted from the central frequency of the radiation $\omega_0$. 
We find the solution of (\ref{eq_four}) for the output field $a_{out,N}(\omega)=S(\omega)a_{in,1}(\omega)$ where TF  is:
\begin{eqnarray}\label{sol_eq}
S(\omega)=e^{i\Phi(\omega)}\prod_{n}\frac{-\frac{k_n}{2}-i(\omega-\Delta_n')+\frac{N_ng_n^2}{{\delta}_n-i(\omega-\tilde{\Delta}_n)}}{\frac{k_n}{2}-i(\omega-\Delta_n')+\frac{N_n g_n^2}{{\delta}_n-i(\omega-\tilde{\Delta}_n)}},
\end{eqnarray}
where $\Phi(\omega)=(\omega+\omega_0)(z_N-z_1)/c$.

By using time-reversal property of photon echo protocols \cite{Moiseev2001,Moiseev2004,Riedmatten2008,Tittel2009,Moiseev2010} we can revise macroscopic spin coherence by rephasing spin states and initiating echo signal emission, respectively.
Following this way and taking into account weak spin decoherence $\gamma\ll\kappa_1$, we get from equation (\ref{eq}), the total QM efficiency  $\eta_{tot}\approx\eta_{stor}^2$, where recording efficiency

\begin{eqnarray}\label{eff}
\eta_{stor}(\omega)= 1-|S(\omega)|^2-\sum_n\frac{\gamma_n\langle b_n^{\dagger}(\omega)b_n(\omega)\rangle}{\langle a_{in,1}^{\dagger}(\omega)a_{in,1}(\omega)\rangle}.
\end{eqnarray}
\noindent
where $\langle...\rangle$ are calculated by using initial field-atoms quantum state $|\Psi_{in}\rangle=|G_s\rangle|\psi_{in,1}\rangle$.
For calculation of optimal system parameters providing maximum quantum efficiency $\eta_{stor}(\omega)$, for simplicity we imply $\gamma_1=...=\gamma_N=\gamma=0$ (where $\eta_{stor}(\omega,\gamma=0)=\eta_{stor}^0(\omega)$).
In this case the upper bound of the efficiency is given by the expression $\eta_{stor}\approx\eta_{stor}^0-\xi\gamma/\kappa_1$ with $\xi\approx1\div10$, where
\begin{eqnarray}\label{eff0}
\eta_{stor}^0(\omega)=1-|S(\omega)|^2,
\end{eqnarray}
actually determines the upper limit of the qualitative capabilities of the studied cascade CFC QM for the storage of input signal field in the several long-lived spin ensembles via the ring high-Q resonators.
Further relaxation during storage time $T$ will be determined by the decoherent processes (with long decay time $T_2\gg 1/\kappa_1$) in the long-lived spin system ($\eta_{tot}^0(\omega,T)\approx e^{-2T/T_2}\eta_{stor}^0(\omega)$ for the input single-photon field). 

\section{Optimization procedure}
In general, TF (\ref{sol_eq}) has very complex spectral properties \cite{Swanson2007,Koziel2011,Tamiazzo2017} and many mathematical aspects of our topic intersect with the fundamental problems of the theory of filters.
However, we show that under the certain conditions, the real system can acquire the spectral properties of TF corresponding to the highly effective sensor \cite{Cheng2014}, broadband filter \cite{Swanson2007,Rosenberg2013}, QM \cite{Perminov2017spectral} or QI.
An ideal broadband QI with an infinite working spectral bandwidth corresponds to $S(\omega)=0$.
Considering the last equality for $S(\omega)$ as an approximate condition for the problem of algebraic polynomial optimization, we have:
$S(0)\rightarrow0,\sum\limits_{m=1}^{N_{opt}}|\textrm{numer}(S(m\tilde{\Delta}_n/(2N_{opt})))|^2\rightarrow \textrm{min},N_{opt}=2N-1$
for $4N+1$ free parameters $\{\tilde{\delta}={\delta}_n,g_n,\Delta_n\}$ (where $\textrm{numer} (S(\omega))$ is the numerator of the rational function $S(\omega)$).

We solved the equality $S(\omega)\approx0$ accurately only for several spectral points.
We note that the first of the conditions of the form $S(0)=0$ is analogous to the standard condition for a impedance matched resonator QM \cite{Moiseev2010,Moiseev_2017_PRA}, which determines the spectral efficiency in the central part of the band for a broadband QM, and additional conditions allow improving the spectral efficiency at edges of the used frequency band.

\begin{figure}[t]
	\includegraphics[width = 0.45\textwidth]{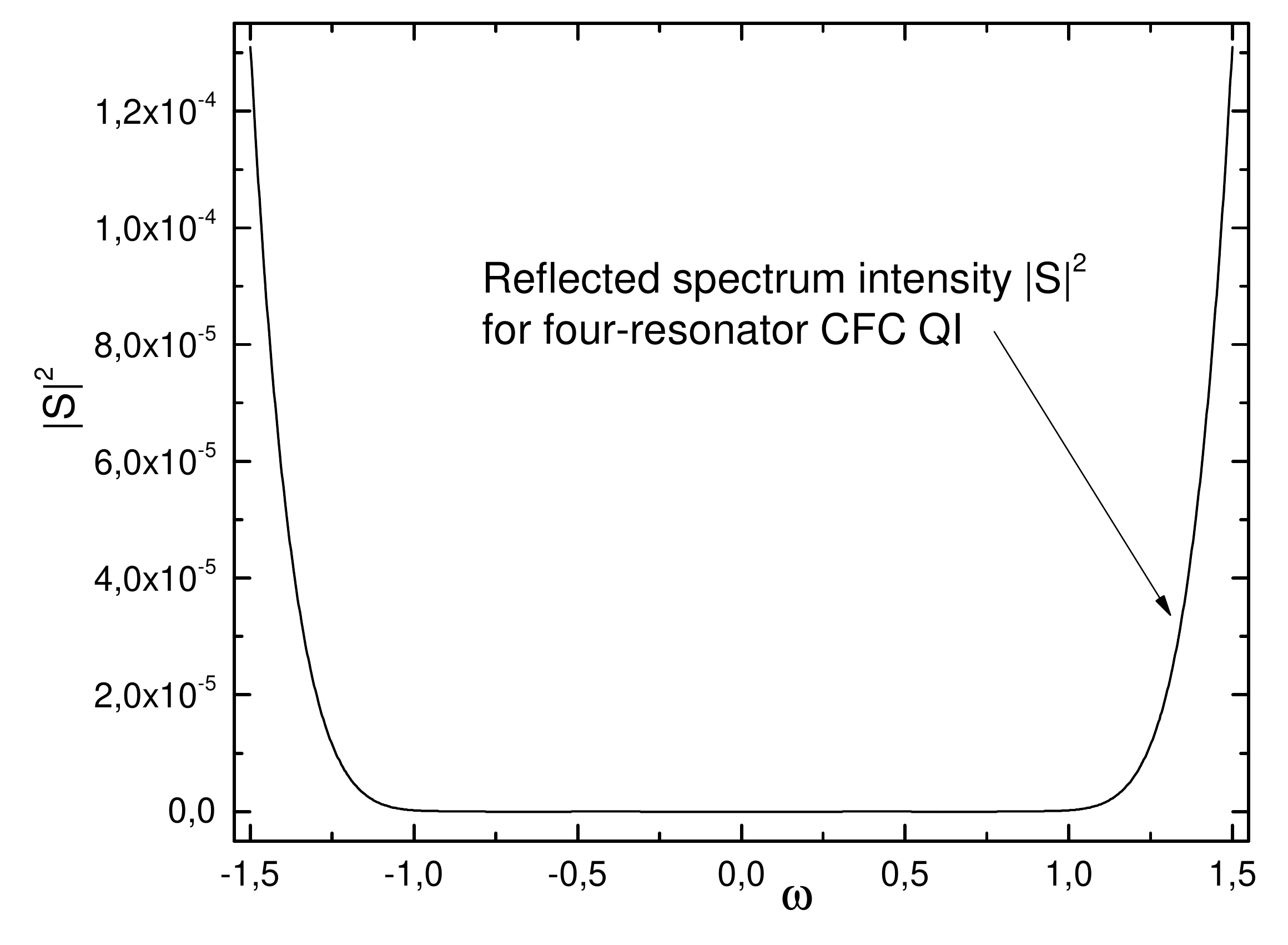}
	\caption{Intensity of the reflected spectrum $|S(\omega)|^2$ for the optimal CFC QI.}
	\label{int_spectra}
\end{figure}

Here we consider the particular case of four resonators ($N=2$) and after numerical calculations (for $\Delta=1$), we obtain a configuration which is a symmetric with respect to the indices
$\tilde{\delta}=1.8,k_{\mp1}=3.27,k_{\mp2}=2.03,N_1^{\frac{1}{2}}g_{\mp1}=1.78,N_2^{\frac{1}{2}}g_{\mp2}=1.49,\Delta_{\mp1}=0.48,\Delta_{\mp2}=1.13$ and the intensity for the reflection spectrum $|S(\omega)|^2$ depicted in figure \ref{int_spectra}.
From where we found the optimal absorption coefficients of the resonator modes by spin ensembles $N_1g_1^2T_2^*=1.76$ and $N_2g_2^2T_2^*=1.23$.
The absorption coefficients are much smaller than the optimal absorption coefficient $\kappa$, which should be realized at the usual impedance matching conditions if the spin system will be placed in one common resonator with spectral width $\kappa\gg\Delta=1$ \cite{Moiseev2010}.
This observation shows that the total number of spins can be reduced in the studied cascade CFC QM scheme.
The spectral behavior of $|S(\omega)|^2$ in figure \ref{int_spectra} demonstrates the high quality of the CFC QI.
$|S(\omega)|^2$ is characterized by an almost rectangular spectral plateau of a fairly homogeneous form in the frequency range $\omega\in[-1.5;1.5]$, where the efficiency $\eta_{stor}^0$ of the absorption of the signal field by the CFC system reaches $99.99$ \%.

\section{Conclusion}
First experimental realization of QI on the system of ring microwave resonators \cite{Petrovnin2018} demonstrated a possibility of broadband storage for small number of resonators. 
In this paper, we theoretically show that optimization of all the parameters in CFC QI with 4 ring resonators is possible for the signal storage on 4 spin ensembles covering a wide frequency band with a quite large efficiency 99.99\%. 
The storage requires using sufficiently high-quality microresonators \cite{Megrant2012,Armani2003,Vahala2003} for which the total losses are limited by the effective decay constant $\sim\gamma/k_1+T/T_2$.

Rather simple optimization scheme used by us at several spectral points does not require large computational resources and it can be also analyzed analytically on the basis of applied methods of algebraic geometry \cite{Amari2006,Shakirov2007,Dolotin2007}.
The studied scheme can be also realized on the system of high-quality WGM (whispering gallery modes) microresonators \cite{Gorodetsky1994,Gorodetsky1999} coupled to the external optical waveguide by including resonant atomic ensembles in each microresonator.
Herein the coherent control of the optical atomic coherence can be carried out by an additional lasers tuned to other eigenfrequencies of WGM microresonators.

In addition, the proposed cascade CFC QI can be used to combine several different types of QM devices (for the systems with different types of inhomogeneous broadening) into a single broadband QM block with a higher quality spectral profile \cite{Perminov2017spectral}.
In the proposed way it is possible to create a superefficient hybrid QM devices \cite{Kurizki2015}, which can be integrated into the quantum computer circuit on the basis of already existing technologies \cite{Brecht_2016,Pierre2014,Du2016,Melloni2010,Romanenko2014,Huet2016,Gu2017}.

\ack
This work was financially supported by a grant of the Government of the Russian Federation, project No. 14.Z50.31.0040, February 17, 2017. 

\section*{References}

\bibliographystyle{iopart-num}
\bibliography{CFC_QI}

\end{document}